\documentstyle[12pt]{article}
\begin{document}

\begin{center}
{\bf Short popular review of quantum electromagnetodynamics} \\

\vspace{0.5cm}
Rainer W. K\"uhne \\
{\em kuehne@theorie.physik.uni-wuppertal.de}
\end{center}

\vspace{1cm}

\noindent
{\bf The aim of this note is to give a short and popular review 
of the ideas which led to my model of magnetic monopoles and 
my prediction of the second kind of electromagnetic radiation.  
I will also point out the many and far-reaching consequences if these 
magnetic photon rays would be confirmed.}

\vspace{1cm}

\noindent
The discovery of a second kind of light would  
be a multi-dimensional scientific revolution. It would shake the foundations 
of modern physics in many ways. It would be experimental 
evidence of physics beyond the standard theory of particle physics. 
The standard theory includes the Weinberg-Salam theory from 1967/1968 
and quantum chromodynamics from 1973. The observation would 
require not only that the theory of quantum electrodynamics formulated 
in 1948/1949 has to be extended. It would challenge also the 
Copenhagen interpretation of quantum mechanics formulated in 
1927/1928. Furthermore, the new kind of light would violate the relativity 
principle of special relativity from 1905 and would require a symmetrization 
of Maxwell's equations from 1873.

The existence of the second kind of light was predicted theoretically. 
It can be understood by the following argumentation.

In 1948/1949 Tomonaga, Schwinger, and Feynman introduced quantum 
electrodynamics. It is the quantum field theory of electric and 
magnetic phenomena. This theory has one shortcoming. 
It cannot explain why electric charge is quantized, i.e. why it 
appears only in discrete units.

In 1931 Dirac introduced the concept of magnetic monopoles. 
He has shown that any theory which includes magnetic monopoles 
requires the quantization of electric charge.

A theory of electric and magnetic phenomena which includes Dirac 
monopoles can be formulated in a manifestly covariant and 
symmetrical way if two four-potentials are used. Cabibbo and Ferrari 
in 1962 were the first to formulate such a theory. Within the 
framework of a quantum field theory one four-potential corresponds 
to Einstein's electric photon from 1905 and the other four-potential 
corresponds to Salam's magnetic photon from 1966.

In 1997 I have shown that the Lorentz force between an electric 
charge and a magnetic charge can be generated as follows \cite{1}. 
An electric charge 
couples via the well-known vector coupling with an electric photon and 
via a new type of tensor coupling, named velocity coupling, with a 
magnetic photon. This velocity coupling requires the existence of a 
velocity operator.

For scattering processes this velocity is the relative velocity 
between the electric charge and the magnetic charge just before 
the scattering. For emission and absorption processes there is no 
possibility of a relative velocity. The velocity is the absolute 
velocity of the electric charge just before the reaction.

The absolute velocity of a terrestrial laboratory was measured by 
the dipole anisotropy of the cosmic microwave background radiation. 
This radiation was detected in 1965 by Penzias and Wilson, its 
dipole anisotropy was detected in 1977 by Smoot, Gorenstein, and Muller. 
The mean value of the laboratory's absolute velocity is 371 km/s. 
It has an annual sinusoidal period because of the Earth's motion 
around the Sun with 30 km/s. It has also a daily sinusoidal period 
because of the Earth's rotation with 0.5 km/s.

According to my model from 1997 \cite{1} each process that produces 
electric photons does create also magnetic photons. The cross-section 
of magnetic photons in a terrestrial laboratory is roughly one 
million times smaller than that of electric photons of the same energy. 
The exact value varies with time and has both the annual and the 
daily period.

As a consequence, magnetic photons are one million times harder to 
create, to shield, and to absorb than electric photons of the same 
energy.

The easiest test to verify/falsify the magnetic photon is to illuminate a 
metal foil of thickness $1,\ldots ,100\mu$m  by a laser beam (or any other 
bright light source) and to place a detector (avalanche diode or 
photomultiplier tube) behind the foil. If a single foil is used, then the 
expected reflection losses are less than 30\%. If a laser beam of the 
visible light is used, then the absorption losses are less than 15\%. My 
model \cite{1} has to be considered as falsified if the detected intensity is 
less than $1.0\times 10^{-12}$ times the intensity that would be detected 
if the metal foil were removed and the laser beam would directly illuminate 
the detector. 

The observation of magnetic photon rays would be a multi-dimensional 
revolution in physics. Its implications would be far-reaching.

(1) The experiment would provide evidence of a second kind of electromagnetic 
radiation. These magnetic photon rays are more penetratable than 
electric photon light of the same wavelength. Hence, they may find 
applications in medicine where X-ray and ultrasonic diagnostics are 
not useful. X-ray examinations include a high risk of radiation damages, 
because the examination of teeth requires high intensities of 
X-rays and genitals are too sensible to radiation damages. Examinations 
of bones and the brain may also become possible.

(2) The experiment would confirm the existence of a new vector gauge boson, 
Salam's magnetic photon from 1966. It has the same quantum numbers as 
Einstein's electric photon, i.e. spin of one, negative parity, zero 
rest mass, and zero charge.

(3) A positive result would provide evidence of an extension of quantum 
electrodynamics which includes a symmetrization of Maxwell's 
equations from 1873.

(4) The experiment would provide indirect evidence of Dirac's magnetic 
monopoles from 1931 and the explanation of the quantization of electric 
charge. This quantization is known since Rutherford's discovery of the 
proton in 1919.

(5) My model describes both an electric current and a magnetic current, 
even in experimental situations which do not include magnetic charges. 
This new magnetic current has a larger specific resistance in conductors 
than the electric current. It may find applications in electronics.

(6) Dirac noticed in 1931 that the coupling constant of magnetic 
monopoles is much greater than unity. This raises new questions 
concerning the perturbation theory, the renormalizability, and the 
unitarity of quantum field theories.

(7) The intensity of the magnetic photon rays should depend on 
the absolute velocity of the laboratory. The existence of the 
absolute velocity would violate Einstein's relativity principle of special 
relativity from 1905. It would be interesting to learn whether 
there exist further effects of absolute motion.

(8) The supposed non-existence of an absolute rest frame was the only 
argument against the existence of a luminiferous aether. If the 
absolute velocity does exist, we have to ask whether aether 
exists and what its nature is.

(9) When in 1925 Heisenberg introduced quantum mechanics, he argued 
that motion does not exist in this theory. This view is taken also 
in the Copenhagen interpretation of quantum mechanics formulated in 
1927/1928 by Heisenberg and Bohr. The appearance of a velocity 
operator in my model challenges this Copenhagen interpretation. 
Mathematically, the introduction of a velocity (and force) operator 
means that quantum mechanics has to be described not only by partial but 
also by ordinary differential equations.

(10) Magnetic photon rays may contribute to our understanding of 
several astrophysical and high energy particle physics phenomena 
where relativistic absolute velocities appear and where electric 
and magnetic photon rays are expected to be created in comparable 
intensities.

(11) Finally, the other interactions may show similar dualities. 
The new dual partners of the known gauge bosons would be the 
magnetic photon, the isomagnetic W- and Z-boson, and the 
chromomagnetic gluons. In 1999 I argued that the dual 
partner of the graviton would be the tordion \cite{2}. This boson has a spin 
of three and is required by Cartan's torsion theory from 1922 which 
is an extension of Einstein's general relativity from 1915.

\end{document}